\documentclass[aps,prl,twocolumn,showpacs,superscriptaddress,groupedaddress]{revtex4}

\usepackage{graphicx}
\usepackage{dcolumn}   
\usepackage{bm,braket}        
\usepackage{amssymb}   
\usepackage{amsmath}
\usepackage{url,color}
\usepackage{layout}
\usepackage{epsfig}
\usepackage{graphicx}
\usepackage{epstopdf}
\usepackage{booktabs}
\usepackage{float,CJK}
\usepackage[hyperindex,breaklinks]{hyperref}

\def\be{\begin{equation}}
\def\ee{\end{equation}}
\def\ba{\begin{eqnarray}}
\def\ea{\end{eqnarray}}

\begin{document}
\title{Quantum independent set problem and non-abelian adiabatic mixing}
\begin{CJK}{UTF8}{gbsn}
\author{Biao Wu(吴飙)}
\email{wubiao@pku.edu.cn}
\affiliation{International Center for Quantum Materials, School of Physics, Peking University, 100871, Beijing, China}
\affiliation{Wilczek Quantum Center, School of Physics and Astronomy, Shanghai Jiao Tong University, Shanghai 200240, China}
\affiliation{Collaborative Innovation Center of Quantum Matter, Beijing 100871,  China}
\author{Hongye Yu(余泓烨)}
\affiliation{International Center for Quantum Materials, School of Physics, Peking University, 100871, Beijing, China}
\author{Frank Wilczek}
\affiliation{Center for Theoretical Physics, MIT, Cambridge MA 02139 USA}
\affiliation{T. D. Lee Institute, Shanghai Jiao Tong University, Shanghai 200240, China}
\affiliation{Wilczek Quantum Center, School of Physics and Astronomy,
Shanghai Jiao Tong University, Shanghai 200240, China}
\affiliation{Department of Physics, Stockholm University, Stockholm SE-106 91 Sweden}
\affiliation{Department of Physics and Origins Project, Arizona State University, Tempe AZ 25287 USA}

\date{\today}
\begin{abstract}
We present an efficient quantum algorithm for some independent set problems in graph theory, based on non-abelian adiabatic mixing. 
We illustrate the performance of our algorithm with analysis and numerical calculations  for two different types of graphs, with the number 
of edges proportional to the number of vertices or its square. The theoretical advantages of our quantum algorithm over classical algorithms are discussed.   
Non-abelian adiabatic mixing can be a general technique to aid exploration in a landscape of near-degenerate ground states. 
\end{abstract}

\maketitle
\end{CJK}
{\it Introduction -} The supremacy  of quantum  computers over classical computers is 
illustrated by many significant algorithms, in particular,  
the Shor algorithm~\cite{Shor} for factorization and the Grover algorithm~\cite{Grover} for search. 
These algorithms are  based on discrete operations orchestrating  simple quantum gates.  
Algorithms of this kind are called 
quantum circuit algorithms~\cite{ChuangBook}.  

In another paradigm of quantum computing, algorithms are implemented  through 
the design of  Hamiltonians.  Here one starts with an easy-to-prepare initial state, 
allows it evolve dynamically, and at some point makes appropriate measurements.  
(Of course, the Hamiltonians should correspond to potentially realizable circuits.)   
Hamiltonian-based quantum algorithms translate programming problems 
into physical problems, which allow one to exploit familiar physical processes to optimize algorithms. 
A Hamiltonian approach to  quantum search was proposed  in 1998~\cite{Farhi1998}, 
and soon extended to more general ``adiabatic'' algorithms ~\cite{Farhi2000}.  

It has been shown that every 
quantum circuit algorithm can be converted into a quantum adiabatic algorithm, 
whose time complexity is polynomially equivalent (and {\it vice versa})  ~\cite{Dam}~\cite{YHW}.  But the continuum 
approach can suggest different methods, such as the non-abelian mixing discussed here, or resonance, as we will describe elsewhere ~\cite{WHW}. 

\begin{figure}[htbp]
	\includegraphics[width = 0.65\linewidth]{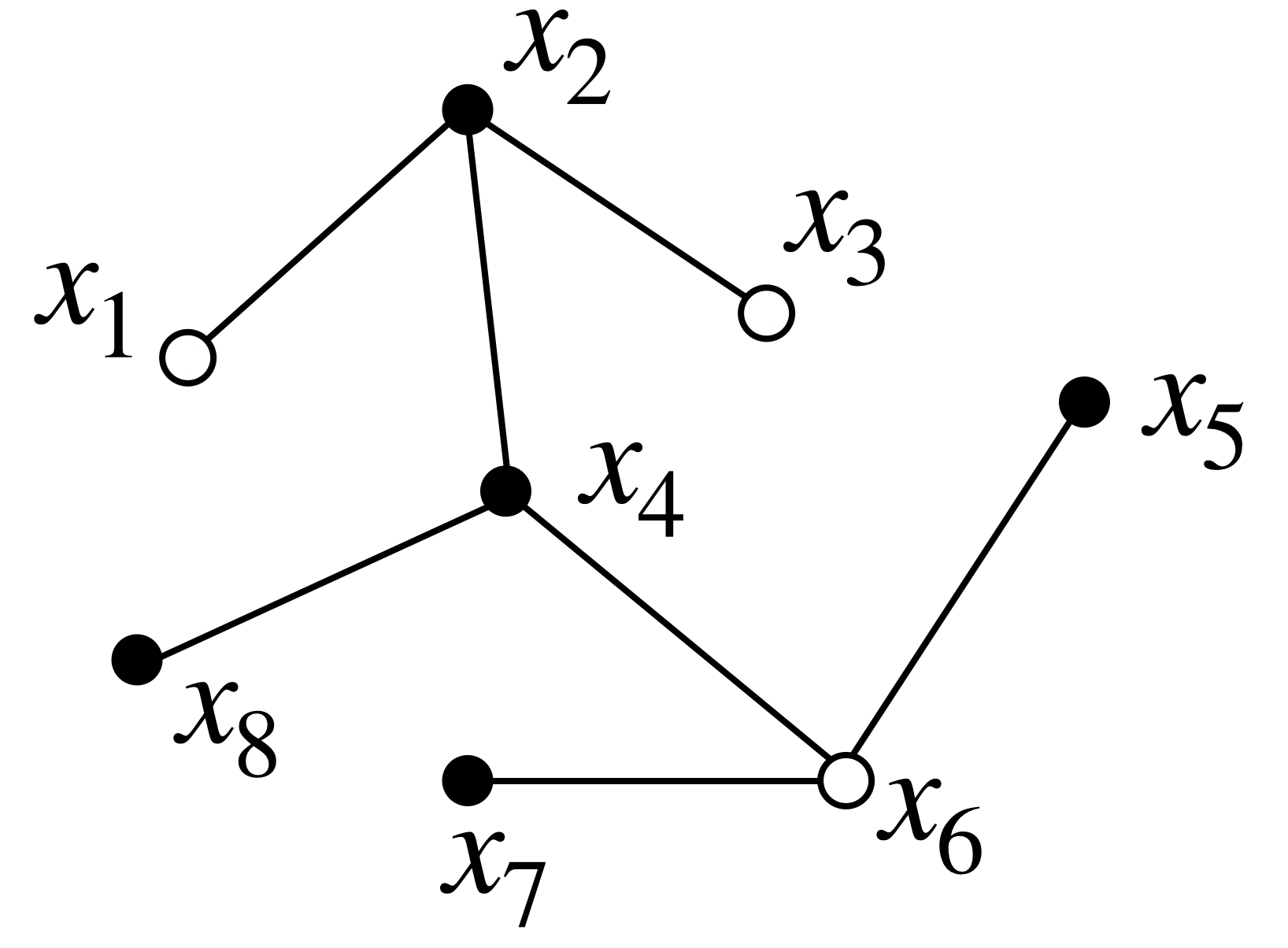}
	\caption{ An independent set of a graph is a set of vertices  no two of which are connected by an edge. 
	Each vertex is assigned a boolean variable: $x_j=1$ if the $j$th vertex is included in an independent set and 
	 $x_j=0$ if not.  For example, the empty circles here form an  independent set that is described by a set of 
	 boolean numbers $(1,0,1,0,0,1,0,0)$. }
	\label{graph}
\end{figure}
Here we present an efficient quantum Hamiltonian algorithm for the independent set problem (see Fig.\ref{graph}).
Any graph has trivial independent sets: the empty set and sets with just one vertex. 
Our aim is to find non-trivial independent sets, with two or, ideally, many more vertices. 
The independent set problem is exactly equivalent to a special class of 2-SAT problem, which
we call all-negated 2-SAT problem. Based on this observation, we are able to 
construct a Hamiltonian  such that its ground states are independent sets of a given graph. We then prepare 
the Hamiltonian system  in one of its trivial ground state, evolve it adiabatically 
along a closed path. This leads to non-abelian adiabatic mixing
in the sub-Hilbert space of degenerate ground states~\cite{WZ} and generates a quantum state
that is roughly an equal-probability superposition of all ground states.  As the
number of non-trivial solutions is much bigger, when we  make a measurement in the end, 
we will likely find a non-trivial solution. 
Numerical results indicate that we are almost certain to find a non-trivial independent set.  
We analyze the performance of our algorithm for two different types of graphs: the number 
of edges proportional to the number of vertices or  its square. 
While finding  solutions to this particular problem is not a pressing issue, our technique brings in some 
physics which is interesting in itself and new in the context of quantum algorithms, and which 
might find more general applications.

{\it Equivalence to 2-SAT -} For a given graph, we can assign a Boolean variable to each 
of its vertices (see Fig.\ref{graph}): $x_j=1$ when the $j$th vertex is chosen for an independent set 
and $x_j=0$ when  it is not. When two vertices $x_i$ and $x_j$ are connected by an edge, it means that $x_i$ and $x_j$ 
can not be simultaneously chosen for one
independent set. This is equivalent to impose the following two-variable clause 
\be
(\lnot x_{i}\lor \lnot x_j)~~~~~~~(i\neq j).
\ee
Therefore, finding an independent set of a graph $n$ vertices with $m$ edges is equivalent to finding a solution to 
a 2-SAT problem which has $n$ variables and whose $m$ clauses are of the above form. Since the clauses involve 
only negated variables, we call it all-negated 2-SAT problem. 
An all-negated 2-SAT problem manifestly has at least $n+1$ 
solutions,  namely $(0,0,0\cdots,0)$ and $n$ assignments that have exactly one 
variable being 1, such as $(1,0,0\cdots,0)$ and $(0,1,0\cdots,0)$. They correspond  to 
the trivial independent sets: the empty set and sets with only one vertex. 
We are interested in finding non-trivial solutions, that is, 
the solutions with at least two 1s. There are generic algorithms of time complexity $O(n)$ to find solutions for 2-SAT problems~\cite{Aspvall,Even}. 
However, these algorithms may well find the trivial solutions.  We need different algorithms to find non-trivial solutions.

{\it Quantum algorithm -} 
For a given graph (or a 2-SAT problem), noticing $x_j=(\hat{\sigma}_j^z+1)/2$, we construct the following Hamiltonian~\cite{Farhi2001}
\be
H_{0}=\Delta\sum_{\langle ij\rangle}(\hat{\sigma}^z_{i}+\hat{\sigma}^z_{j}+\hat{\sigma}^z_{i}\hat{\sigma}^z_{j})\,,
\ee
where the summation $\langle ij\rangle$ is over all edges (or clauses) . All the independent sets  
are the ground states of $H_0$ and vice versa.   The energy gap between the ground states and 
the first excited states is $4\Delta$.

\def\ur{u_{\vec{r}}}
\def\dr{d_{\vec{r}}}

We rotate spin $\hat{\sigma}_j^z$  to an arbitrary direction 
$\vec{r}=\{\sin\bar{\theta}\cos\bar{\varphi},\sin\bar{\theta}\sin\bar{\varphi},\cos\bar{\theta}\}$, and obtain
new spin operator $\hat{\tau}_j=V_j \hat{\sigma}_j^zV_j^{-1}$ with 
\be
V_j=
\begin{pmatrix}\cos\frac{\bar{\theta}}{2} & e^{-i\bar{\varphi}}\sin\frac{\bar{\theta}}{2}\\ 
e^{i\bar{\varphi}}\sin\frac{\bar{\theta}}{2}&-\cos\frac{\bar{\theta}}{2}
\end{pmatrix}=V_j^{-1}\,.
\label{matrix}
\ee
If $\ket{u}_j$ and $\ket{d}_j$ are eigenstates 
of $\hat{\sigma}_j^z$,  that is, $\hat{\sigma}_j^z\ket{u}_j=\ket{u}_j$ and  $\hat{\sigma}_j^z\ket{d}_j=-\ket{d}_j$, 
the eigenstates of $\hat{\tau}_j$ are 
\ba
\label{npm}
\ket{\ur}_j&=&\cos\frac{\bar{\theta}}{2}\ket{u}_j+\sin\frac{\bar{\theta}}{2}e^{i\bar{\varphi}}\ket{d}_j~,\\
\ket{\dr}_j&=&\sin\frac{\bar{\theta}}{2}\ket{u}_j-\cos\frac{\bar{\theta}}{2}e^{i\bar{\varphi}}\ket{d}_j~.
\ea
With $U=V_1\otimes V_2\otimes \cdots \otimes V_n$, 
we can rotate all the spins to the same direction and construct a new Hamiltonian
\be
H_\tau=UH_{0}U^{-1}=\Delta\sum_{\langle ij\rangle}(\hat{\tau}_{i}+\hat{\tau}_{j}+\hat{\tau}_{i}\hat{\tau}_{j})
\ee
It is clear that $H_\tau$ has  the same set of eigenvalues as $H_0$.  The eigenstates of $H_\tau$ can be
obtained  by rotating the ones  of  $H_0$, and have the following form
\ba
\ket{E_\alpha}&=&\ket{\ur}_1\otimes \ket{\dr}_2 \otimes \cdots\otimes\ket{\ur}_{j}\otimes\cdots \otimes
\ket{\ur}_{n}\nonumber\\
&=&\ket{\ur,\dr,\cdots,\ur,\cdots,\ur}\,.
\ea

\begin{figure}[!t]
	    \centering
	    \includegraphics[width=5.0cm]{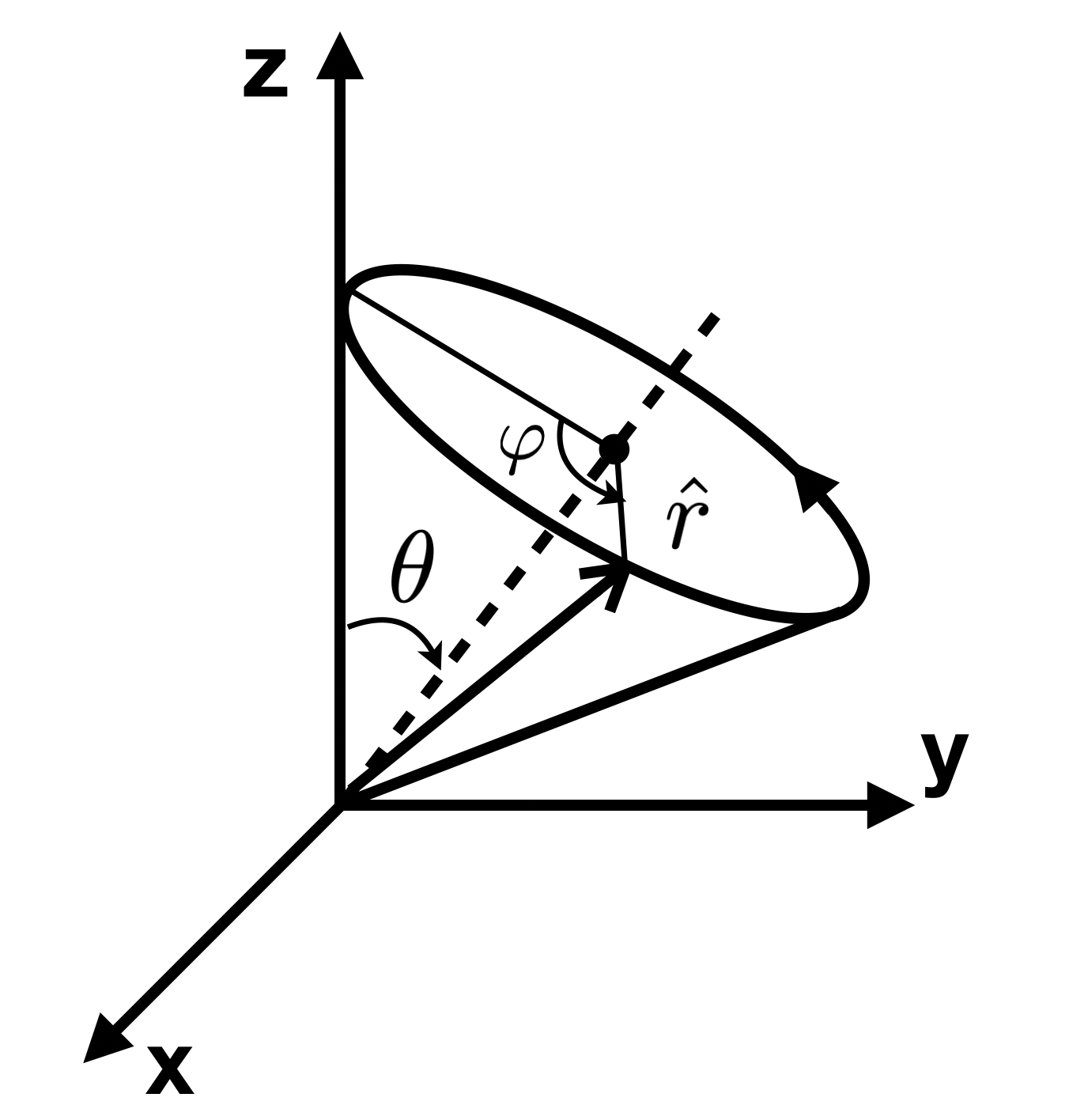}
    	\caption{Adiabatic path in the algorithm. $\theta$ is the angle between 
	the rotating axis and the $z$ axis and $\varphi$ is the angle rotated from the initial direction. Note 
	that $\theta$ and $\varphi$ here are related to but different from $\bar{\theta}$ and $\bar{\varphi}$  in Eq.(\ref{matrix}).}
	\label{loop}
\end{figure}

The Hamiltonian $H_\tau$ is parameterized by the direction $\vec{r}$. With this in mind  
 we propose the following quantum algorithm for the independent set problem: 
\begin{enumerate}
\item prepare the system  at  state $\{-1,-1,\cdots,-1\}$, which corresponds to the empty set $(0,0,\cdots,0)$;
\item set $\vec{r}$ initially along the $z$ axis and slowly change $H_\tau$ by changing  $\vec{r}$ 
along a closed path shown in  Fig.\ref{loop}; 
\item make a measurement after $\vec{r}$ returns to the $z$ direction.
\end{enumerate}

Note that the energy gap $4\Delta$ of $H_\tau$  does not change with $\vec{r}$ and 
is independent of the system size $n$.  Therefore,  the evolution in the above algorithm can be 
made adiabatic by changing $\vec{r}$ with a slow but constant rate. 
As $\{-1,-1,\cdots,-1\}$ is a ground state of $H_{0}$, when $\vec{r}$ changes slowly, 
the system will stay in the sub-Hilbert space spanned by the ground states of $H_\tau$.  
This kind of adiabatic evolution in a sub-Hilbert space
of degenerate eigenstates was studied in Ref.\cite{WZ}, where it is found that an adiabatic evolution along a closed path 
is given by 
\be
W=P\exp i\oint A(t)dt
\ee
where $A$ is the gauge matrix given by $A_{\alpha,\beta}=i\bra{E_\alpha}\partial_t \ket{E_\beta}$ 
and $P$ denotes path ordering. Note that such an adiabatic evolution  of degenerate eigenstates was proposed
to construct quantum gates~\cite{Zanardi}.

We find that $\bra{E_\alpha}\partial_t \ket{E_\beta}$ is not zero only 
when $\ket{E_\alpha}$ and $\ket{E_\beta}$ differ by at most one qubit state.  When $\alpha\neq\beta$, we have 
\be
A_{\alpha, \beta}=i\bra{E_\alpha}\partial_t \ket{E_\beta}=i\bra{\ur}\partial_t \ket{\dr}=\frac{\sin\theta}{2}\frac{d\varphi}{dt}\,,
\ee
where $\theta$ is the angle between the rotating axis and the $z$ axis and $\varphi$ is the rotating angle (see Fig.\ref{loop}). 
When $\alpha=\beta$ and $\ket{E_\alpha}$ has $k$ qubits in state $\ket{\ur}$ and $n-k$ qubits in state $\ket{\dr}$, 
we have 
\ba
A_{\alpha, \alpha}&=&i\bra{E_\alpha}\partial_t \ket{E_\alpha}\nonumber\\
&=&-\big\{k\sin^2\frac\theta2+(n-k)\cos^2\frac\theta2\big\}\frac{d\varphi}{dt}\,.
\ea
Let $A=\widetilde{A}\frac{d\varphi}{dt}$ and we have
\be
W=P\exp \oint i\widetilde{A}(\theta)d\varphi=\exp \Big[2\pi i\widetilde{A}(\theta)\Big]\,,
\label{one}
\ee
where the gauge matrix $\widetilde{A}$ is real and independent of time.  

As the gauge matrix $\widetilde{A}$ has many off-diagonal terms,
it generates a mixing in the sub-Hilbert space of the ground states, producing a quantum state 
that is roughly an equal-probability superposition of all the ground states. When a measurement is made 
at the end of the algorithm, we will likely find a non-trivial ground state since the number of non-trivial solutions
is much bigger than the trivial solutions. 
To illustrate the efficiency of our algorithm, we consider two typical cases:  the number of edges is proportional to 
(I) the number of vertices; (II) the square of 
the number of vertices. 

\begin{figure}[htb]
	\includegraphics[width = 1.0\linewidth]{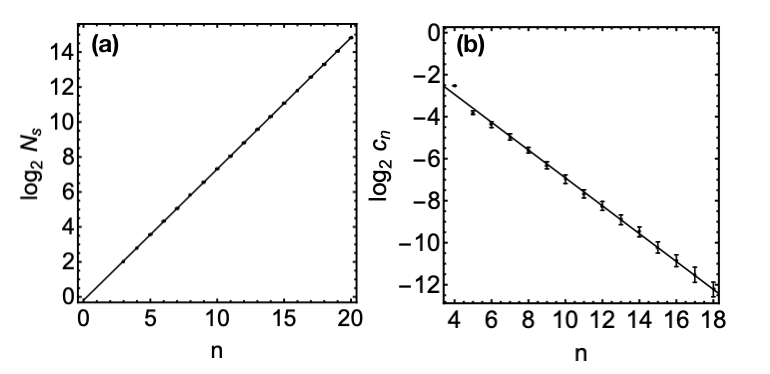}
	\caption{ (a) The number of  independent sets of  a graph  as a function of 
	the number of vertices $n$ for the case  $m=n$.   
	The fitting line is given by $\log_2N_s=0.029 + 0.748n$. 
	 The result is averaged over 1000 instances randomly sampled out of all possible  
	 configurations of edges; the standard error of every data point is around $10^{-3}$ . 
	 (b) The averaged probability  $c_n$ of the $n+1$ trivial solutions in the final state
	 as a function of   $n$.   
	The fitting line is given by $\log_2c_n = -0.444- 0.654 n $. }
	\label{mn}
\end{figure}

{\bf Case I -}  To be specific,  we choose $m=n$. Let $N_s(n)$ be the number of 
all the independent sets of  a given graph.  Our numerical results in Fig.\ref{mn}(a) show that  $N_s$
grows exponentially with  $n$. The fitting gives us $N_s(n) \approx 1.02\times2^{0.748n}$. 
This means that the $n+1$ trivial sets are only a tiny part of all the independent sets when $n$ is large. 

For our quantum algorithm, for simplicity we choose $\theta=\pi/2$, where the 
gauge matrix $\widetilde{A}$ has the simplest form.  We numerically compute 
\be
\ket{\psi_1}=W\ket{\psi_0}
\label{adiab}
\ee
where $\ket{\psi_0}=\{-1,-1,\cdots,-1\}$ is the initial state.  Let $d_n$ be the probability of the $n+1$ trivial solutions 
in the final state $\ket{\psi_1}$ and $c_n=d_n/(n+1)$ be the averaged probability. Our numerical results are 
plotted in Fig.\ref{mn}, where we see $c_n$ decreases exponentially with $n$. Numerical fitting indicates 
$c_n\approx 0.735\times 2^{-0.654n}$. Therefore, we are almost certain to find a non-trivial solution at the end of the algorithm.  
As the gap $4\Delta$ is independent of the problem size $n$,
the time that our adiabatic evolution takes to traverse one loop in Fig.\ref{loop} is  independent of $n$. 
Thus the time complexity of
our quantum algorithm is $O(1)$, and for large $n$ it produces a non-trivial solution with near certainty.



\begin{figure}[h]
	\centering
	\includegraphics[width=0.5\textwidth]{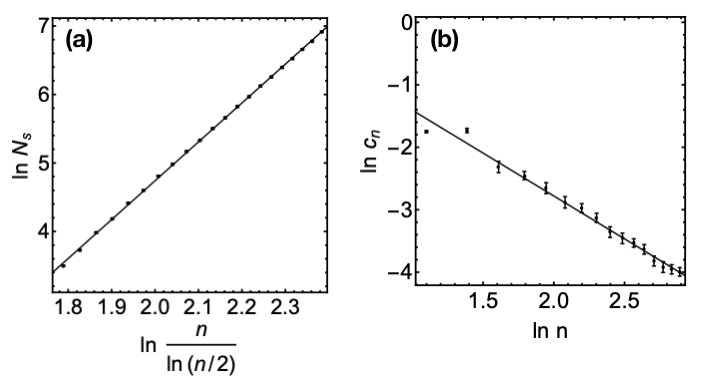}
	\caption{(a) Number of independent sets for graph $m=\lfloor n^2/4\rfloor$ as a function of $n$. 
	 The fitting line is given by $\ln N_s=-6.61131+5.67589 \ln\frac{n}{\ln(n/2)}$. 
	 The result is averaged over 1000 instances randomly sampled out of all possible  
	 configurations of edges; the standard error of every data point is around $10^{-3}$ . 
	 (b) The averaged probability  $c_n$ of the $n+1$ trivial solutions in the final state $\ket{\psi_1}$
	 as a function of  $n$. The fitting line is $\ln c_n=-0.0334482-1.3734 \ln n$.  }
	\label{Sn2}
\end{figure}

{\bf Case II -} We choose specifically $m=\lfloor n^2/4\rfloor$. According to Ref. \cite{frieze1990independence}, 
for such a graph, there exists with almost certainty a maximum independent set of the following size 
\be
k=4\Big(\ln\frac{n}{4\ln(n/2)}+1\Big)
\ee
Since all its subsets are also independent sets,  the number of independent sets $N_s$ is at least 
$
N_s\gtrsim O((n/\ln n)^{4\ln 2})
$. The numerical results  in Fig.\ref{Sn2}(a) show that 
\be
N_s\propto O((n/ \ln n)^{5.7})\,.
\ee

\begin{figure}[t]
	    \centering
	    \includegraphics[width=7cm]{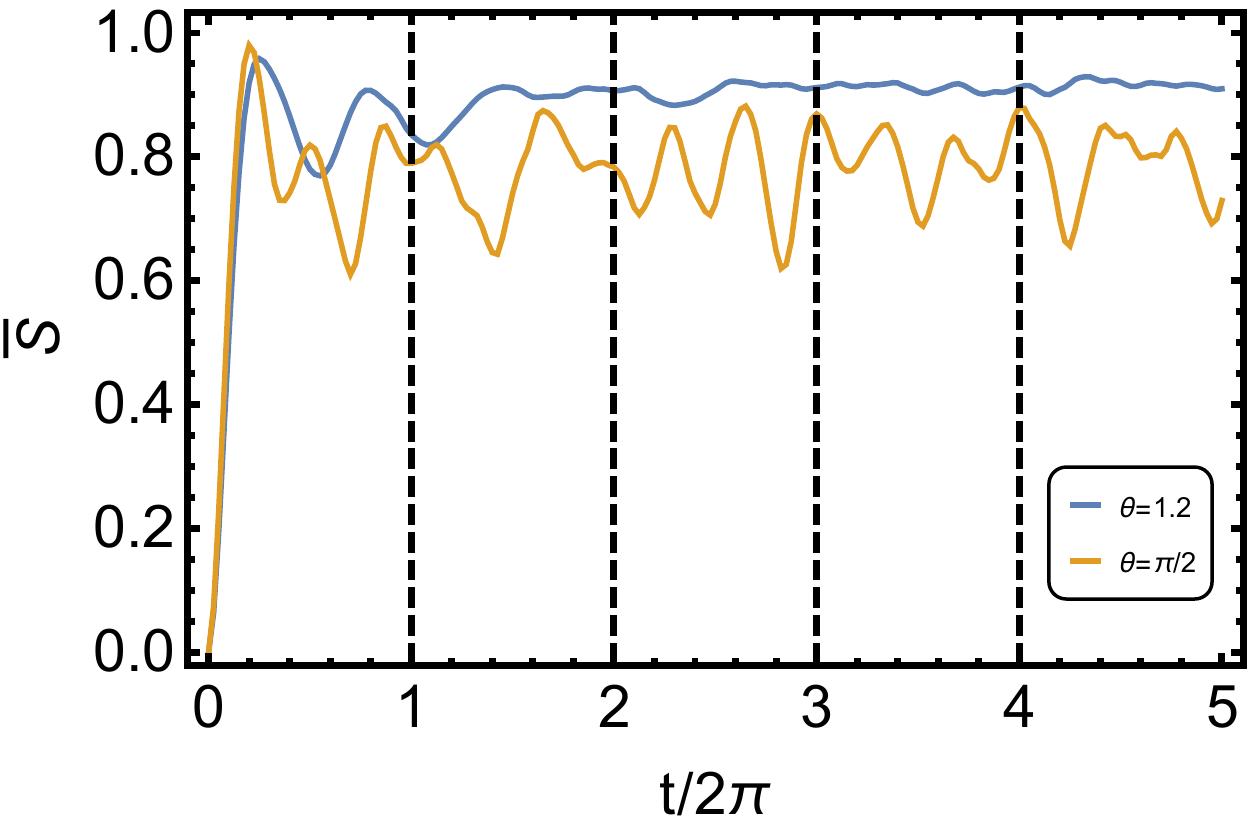}    
	    	\caption{(color online) The time evolution of entropy  $\overline{S}$ for the quantum diffusion in the median graph.  
		The orange line is for a typical independent set with $\theta=\pi/2$; 
		the blue line is for a different typical independent set with $\theta=1.2$.  The averaged or equilibrium
		value of the entropy is $S\approx 0.77$ for $\theta=\pi/2$ and $S\approx 0.88$ for $\theta=1.2$.
		$n=m=12$.}
	\label{entropy}
\end{figure}

For this case,  we  evolve the system along the loop in Fig.\ref{loop} with $\theta=1.2$
to make all possible ground states more evenly distributed in the final quantum state (see later discussion with Fig. \ref{entropy}). 
Our numerical results in Fig.\ref{mn}(b) show that the averaged probability  of finding trivial solutions 
$c_n\propto 1/n^{1.37}$.

We consider  two different  classical algorithms and compare them to our quantum algorithm. 
The first is the generic algorithms for 2-SAT problems~\cite{Aspvall,Even}. With slight modification,
one can expect these algorithms to produce non-trivial solutions with certainty with 
the time complexity of  $O(n)$.  In the second algorithm, one simply picks
up two variables and set them to 1.  For the graph with $m=n$, 
the chance of this randomly-picked solution being wrong
is proportional to $2m/n(n-1)\sim 1/n$, which decreases polynomially with the graph size  $n$. 
In comparison, in our quantum algorithm, the chance of being wrong is exponentially small. 
For the graph with $m=\lfloor n^2/4\rfloor$, the chance of this randomly-picked solution being wrong
is about $n^2/2n(n-1)\sim 1/2$, which is independent of  $n$. In comparison, 
in our quantum algorithm, the chance of producing  trivial solutions decreases 
polynomially with $n$.  If one randomly picks more than two 1s, the chance of being wrong is much greater.  
It is clear that our quantum algorithm holds advantages over both classical algorithms.

\begin{figure}[!ht]
	    \centering
	    \includegraphics[width=5cm]{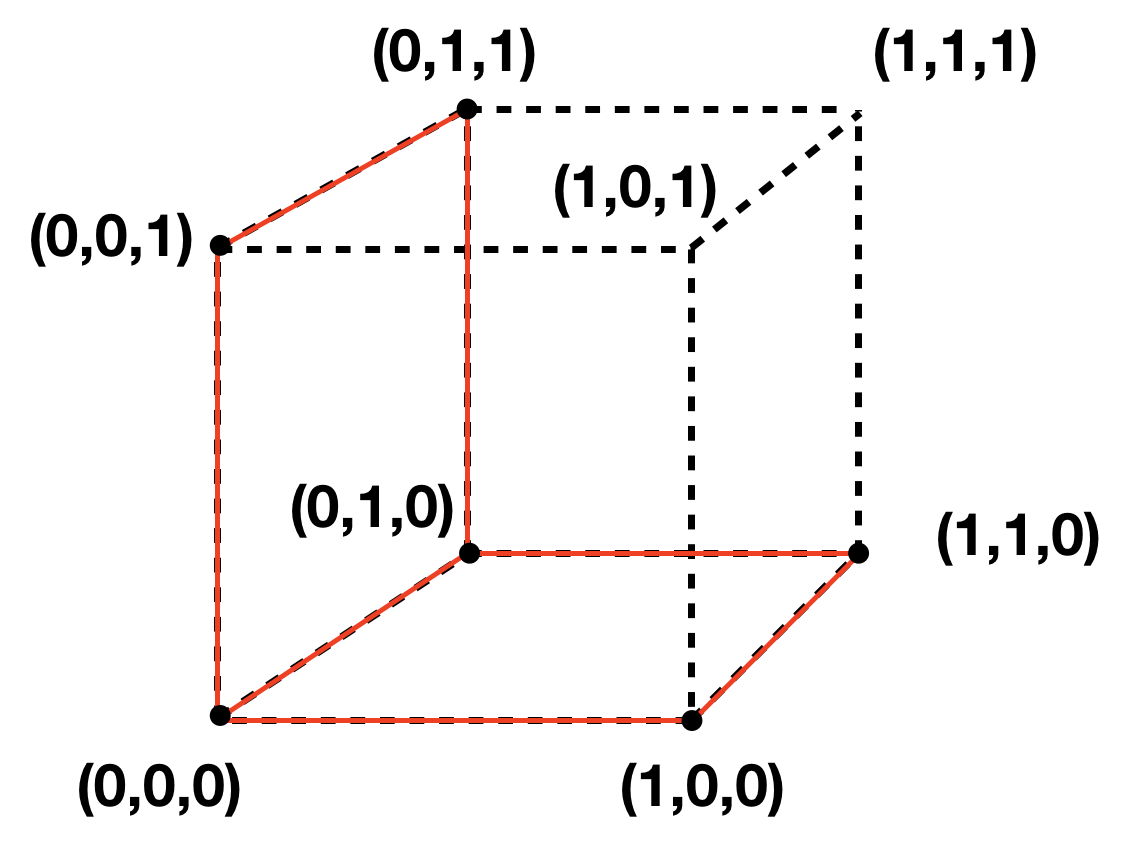}
    	\caption{(color online) A median graph embedded in a cube. Each point represents an independent set  and 
	the red line connects a pair of   independent sets that differ by only one element. }
	\label{cube}
\end{figure}

{\it Quantum diffusion in median graph -} Our algorithm centers on the quantum non-abelian adiabatic mixing
in  a sub-Hilbert space of degenerate ground states. We find that 
such a dynamics process  can also be viewed as a quantum diffusion  in a median graph
which can be embedded in an $n$-dimensional cube (see Fig.\ref{cube}). 

As the solutions of an all-negated 2-SAT problem form a median graph~\cite{Schaefer,Chepoi,wiki2sat}, 
all the independent sets of a graph form a median graph:
each independent set is represented by a point, and a pair of points are connected by a line when the two
independent sets differ  by only one vertex.  This median graph can be embedded in an $n$-dimensional 
cube, as shown in  Fig.\ref{cube} for  $n=3$.  Our Hermitian gauge matrix 
$\widetilde{A}(\theta)$ can be regarded as a Hamiltonian defined on this  median graph: 
the onsite energy is  $\widetilde{A}_{\alpha,\alpha}$
while  off-diagonal element $\widetilde{A}_{\alpha,\beta}$ gives the hopping amplitude  between 
two points $\alpha$ and $\beta$. If we start with an initial wave function localized at $(0,0,\cdots,0)$, this wave function
will spread in the graph and the diffusion process is given by 
\be
\ket{\psi(t)}=\exp \Big[ it\widetilde{A}(\theta)\Big]\ket{\psi_0}\,.
\label{wevo}
\ee
When $t=2\pi$, we recover the adiabatic mixing  in Eq.(\ref{adiab}). So,  the adiabatic evolution in Fig.\ref{loop} 
is just a special case of  quantum diffusion in a median graph  for $t=2n_l\pi$ ($n_l$ is a positive integer).

Let us expand $\ket{\psi(t)}$ in terms of all the solutions
\be
\ket{\psi(t)}=\sum_j a_j(t) \ket{s_j}\,,
\ee
where $\ket{s_j}$ is the $j$th solution. To characterize how widely the wave function is diffused over the median graph, 
we define a quantum entropy 
\be
S(t)=-\sum_j  |a_j(t)|^2\ln |a_j(t)|^2\,.
\ee
It is  called generalized Wigner-von Neumann entropy in Ref.~\cite{gwvn}.  
It is clear that the maximum of $S(t)$ is $\ln N_s$. We define $\overline{S}=S/(\ln N_s)$ and plot $\overline{S}$ 
as a function of $t$ in  Fig.\ref{entropy}. We again consider first the special case $\theta=\pi/2$ (orange line in Fig.\ref{entropy}). 
We observe an interesting behavior of $\overline{S}$: it starts at zero, quickly
rises up to a value very close to one, and eventually oscillates around an equilibrium value. At $t=2\pi,4\pi,6\pi,\cdots$, which 
correspond to adiabatically evolving along the loop in Fig.\ref{loop} one, two, three, $\cdots$ rounds, 
we have $\overline{S}\approx 0.75$. This means that the probability is roughly even distributed among all possible solutions. 
We checked numerically how probability is distributed among different sets of the solutions. For example, 
if the number of solutions with three 1s is $N_3$, then the probability of $\ket{\psi_1}$ in these solutions 
is approximately $N_3/N_s$.

We can reduce the fluctuations of $\overline{S}$ and raise its equilibrium value by choosing a different $\theta$. 
In  Fig.\ref{entropy}, we have  plotted $\overline{S}$ for $\theta=1.2$ (blue line). We see much smaller oscillations 
around a larger equilibrium value. At $t=2\pi,4\pi,6\pi,\cdots$, we have $\overline{S}\sim 0.85$. 

The behavior of $\overline{S}$ in Fig.\ref{entropy} resembles
how a similar quantum entropy behaves in quantum chaotic 
systems~\cite{Neumann1929,von2010proof,han2015entropy,gwvn}: rises up rapidly from a low initial value 
and quickly settles into an equilibrium value. 
By comparing the two lines in Fig.\ref{entropy},  we see that when $\theta$ deviates from the special 
value $\pi/2$,  $\widetilde{A}(\theta)$ tends to be more chaotic. 


{\it Applications and perspective -} The key of our  algorithm, adiabatic non-abelian mixing,  can be applied to other problems 
that have multiple solutions with  one or more solutions easy to find or already found. For example, 
a class of quantum 2-SAT problems have multiple solutions and one of their trivial solutions
is precisely $\ket{-1,-1,-1,\cdots,-1}$~\cite{bravyi,Beaudrap,Arad,Farhi2016,debeaudrap}.

The maximum independent set problem for a graph is a NP-hard problem. Our analysis in the above 
puts this problem in a new perspective. The maximum independent set corresponds to 
the point which is farthest from the original point $(0,0,\cdots,0)$. In our algorithm, 
a quantum particle originally at $(0,0,\cdots,0)$ will indeed arrive at this farthest point through quantum 
diffusion, but with very small probability.  Our understanding of quantum diffusion may help us to find a way to increase 
this probability significantly. 


In sum, we have presented an efficient quantum algorithm for independent set problems 
which exploits the  non-abelian adiabatic mixing in a sub-Hilbert space of degenerate eigenstates. 
According to our numerical results, the time complexity of our algorithm is of $O(1)$, and 
holds advantages over classical algorithms. 

B.W. and H.Y. are  supported by the The National Key R\&D Program of China (Grants No.~2017YFA0303302, No.~2018YFA0305602). F.W. is 
supported by the Swedish Research Council under Contract No. 335-2014-7424. In addition, 
U.S. Department of Energy under grant Contract No.de-sc0012567,  and by the European Research Council under grant 742104.


\end{document}